# Harmonization of Noise Measurement Methods: Measurements of radio impulsive noise from a specific source

Marta Fernández, Iratxe Landa, *Member,* Amaia Arrinda, *Senior Member, IEEE*, Rubén Torre and Manuel Velez, *Member, IEEE*

All the authors are with the Communications Engineering Department of the University of the Basque Country, Faculty of Engineering, Alameda Urquijo s/n. 48013, Bilbao, Spain.

**This article describes a procedure for measuring and evaluating radio impulsive noise (IN) from a specific source. A good knowledge of the noise caused by different sources is essential to plan radio services and to ensure good quality of service. Moreover, it is necessary to harmonize the noise measurement methods in order to achieve results that can be mutually compared. This paper not only provides the steps that should be followed to make proper measurements, but also specifies the appropriate parameters to characterize the impulsive noise when it is generated by a principal source. A detailed description of parameter calculation is presented, based on the recommendations of the International Telecommunication Union (ITU). This work gives answer to the request made in ITU-R 214-4/3 question, which suggests determining the appropriate parameters to describe the noise when it has an impulsive characteristic.**

## 1. Introduction

The evaluation of the radio noise levels is crucial to ensure an efficient spectrum management. Even though several measurement campaigns have been conducted, the continuation of the radio noise study is required, since noise levels have obviously changed in the last years. This has happened because of the evolution of technology, especially in areas with high human activity where noise sources have increased. Furthermore, in the coming years, new ways of communication which incorporate sensors into everyday ordinary objects (Internet of Things) will be developed. So, the services will require more accurate radio planning.

Radio noise comprises two main components: White Gaussian Noise (WGN) and Impulsive Noise (IN). Impulsive noise is an important aspect of man-made noise in particular and can be considered as emissions that are present only for a certain percentage of the time, usually consisting of pulse trains of a limited, short duration and sometimes repeating at a certain rate [1]. These short pulses can be integrated into an event called burst.

When studying impulsive noise, fast data sampling is required and data evaluation is more complex than when analyzing Gaussian noise. Besides evaluating noise levels, it is important to study the noise sources. Therefore, this work provides a methodology for taking measurements and processing them when the noise is generated by a main impulsive noise source. As a result, the values of the parameters that characterize IN are obtained. One of the benefits of following this procedure is that the final results can be easily compared, so the effect of the different sources can be also evaluated.

This study has been done following the ITU-R Recommendation SM.1753 [2]. However, some contributions have been added, especially because this recommendation is not focused on analysing a specific impulsive noise source. After this study it is concluded that some of the recommended parameters in [2] are not suitable for characterizing impulsive noise from some specific sources. This is the case of burst repetition period. In addition, in this study two new parameters are recommended: the number of bursts and the separation between bursts.

Impulsive noise sources have been identified in the literature and different methods have been used to characterize them. Examples of impulsive noise sources previously studied include fluorescent tubes [3,4], photocopiers, elevators [3,5], microwave ovens [5], file shredders [6] or vehicle ignitions [7]. There are also studies that try to model the effect of specific impulsive noise sources on power line communications [8-10]. The possibility of impulsive noise rejection by polarization orthogonality has also been studied [11].

Section 2 explains the appropriate parameters to characterize radio noise, when studying impulsive noise generated by a main source. Section 3 describes the different types of measurements required to achieve the objective of this work. In Section 4, the procedure for taking measurements is defined. Moreover, some specifications about the equipment and the environment are given. Section 5 and Section 6 describe the methods for processing the data collected and the result presentation, respectively. Finally, the discussion and conclusions are presented in Sections 7 and 8.

## 2. Impulsive Noise Characterization

When studying impulsive noise, first it is necessary to characterize the White Gaussian noise present in the location where the IN measurements will be taken. Hence, in this Section the parameters to describe both components are presented.

To characterize White Gaussian noise it is sufficient to know the r.m.s. (root mean square) noise level. However Impulsive noise component is much more difficult to characterize, since the parameters that describe impulsive noise cannot be measured directly. Instead, these parameters are later determined in the data processing. The measurement equipment has to collect samples at a very high speed in order to allow obtaining the following impulsive noise parameters [1,2].

- Number of bursts
- Burst level or amplitude
- Burst length or duration
- Burst separation

These are the appropriate parameters to describe the IN generated by a main source. In these cases, impulses are usually associated to changes in the operation of the device that are called events.

An example of the bursts obtained during a measurement after data processing is plotted in Figure 1. In this figure three bursts are present, and the parameters to describe each burst are given: burst duration and burst amplitude. Moreover, the separation between bursts is represented, since it is important when more than one burst appear. It is necessary to calculate average values of these parameters, as is explained in Section 5, in order to characterize the impulsive noise present in the measurement.

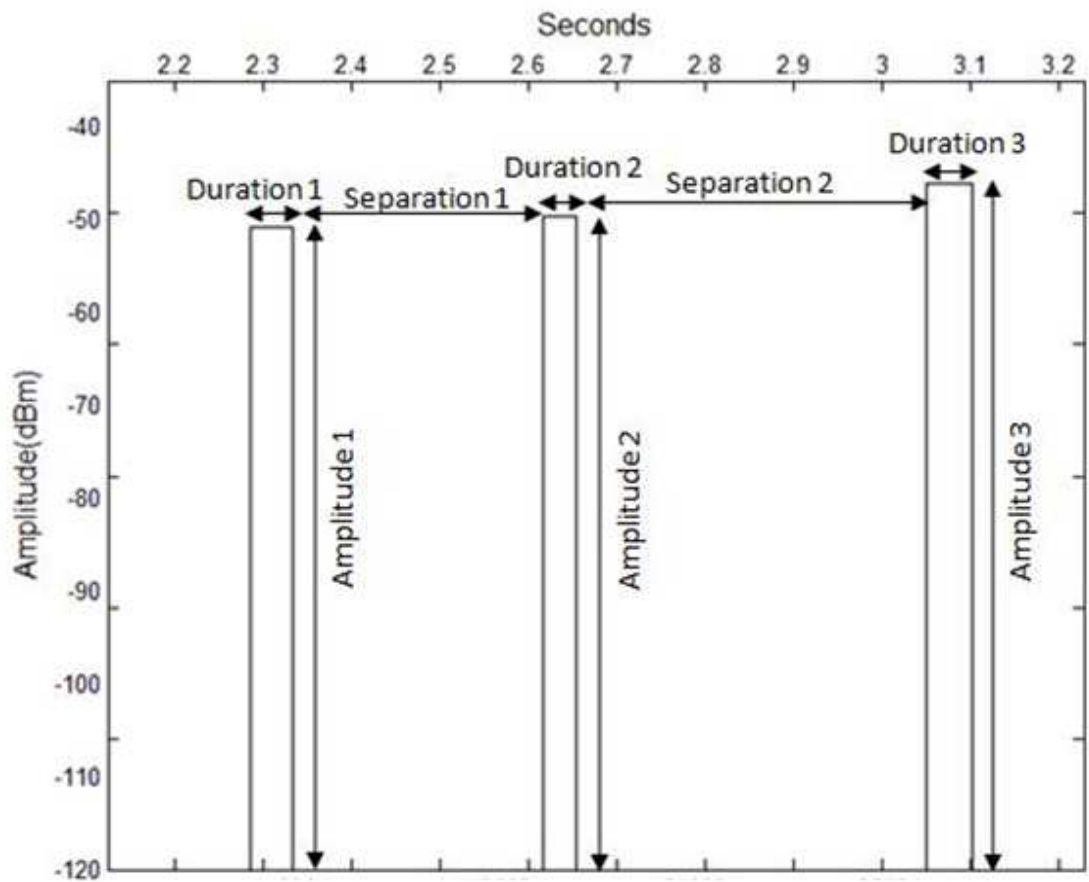

Figure 1. Burst parameters.

## 3. Measurement Types

Two different types of measurements are essential for analyzing an impulsive noise source.
- WGN measurement: a measurement when the source of interest is not working in order to determine the level of WGN. In each location, a measurement per frequency is required.
- IN measurements: the measurements when the specific source is working, so pulses can be detected.

Making a comparison between both types of measurements, impulsive noise generated by a source can be determined

## 4. Measurement Process

One of the purposes of this study is to establish a procedure to measure impulsive noise generated by a main source. Apart from obtaining measurements that allow the study of impulsive noise, the methodology makes it possible to compare different results, even when different sources or various frequencies are studied.

Some requirements must be fulfilled in order to achieve satisfactory results.

**4.1 Equipment Specifications**

The basic equipment needed for impulsive noise measurements consists of a receiver, an antenna and a computer running the control program. An example of the equipment used to take the measurements is shown in Figure 2.

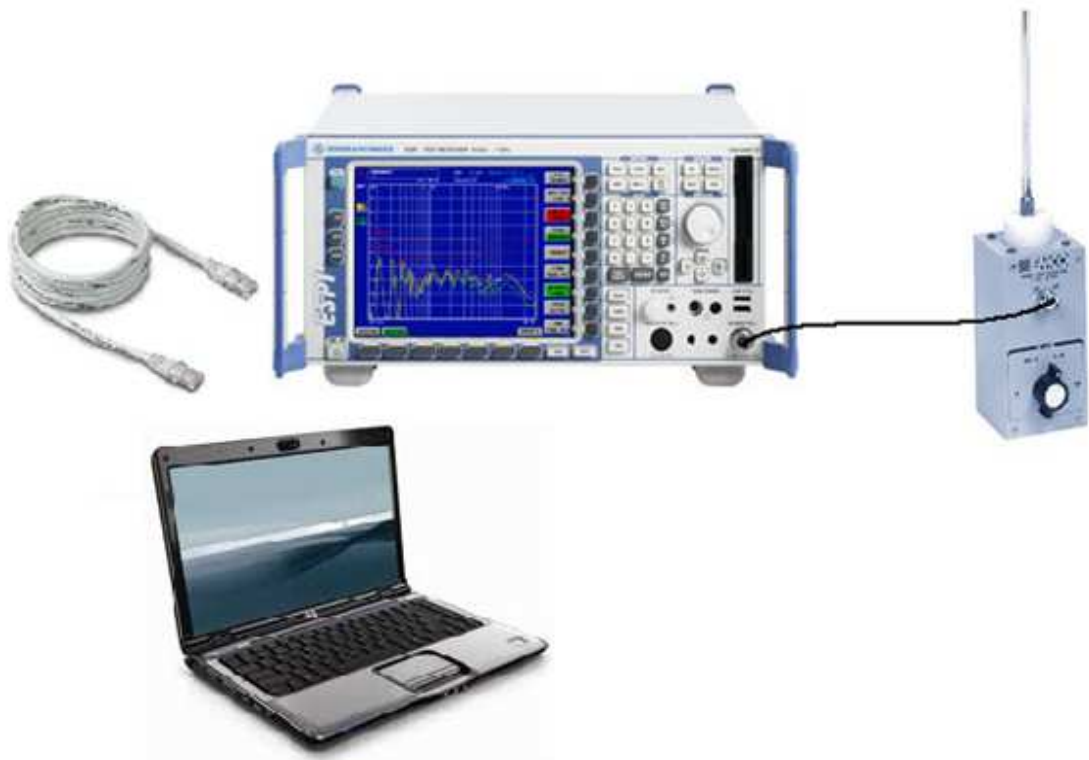

Figure 2. Measurement equipment.

A professional test receiver is recommended for noise measurements. The internal noise should be lower than the external noise level. Further details about the receiver are available in [2]. The recommended antenna is a broadband passive non-directional and sensitive antenna. The Recommendations ITU-R P.372 [12] and ITU-R SM.1753 [2] explain the antenna specifications to measure noise.

In this study, the receiver used is EMI ESPI3 of Rohde & Schwarz. The program to control the equipment and capture data has been designed using Matlab. The receiver manufacturer provides the required commands and the software interface libraries [13]. The control program can be developed using a programming language such as Matlab or C.

The ETS-Lidgren 3303 rod antenna has been used. This is a passive broadband electric field monopole antenna with a frequency range of 1 kHz to 30 MHz.

The software running on the computer sets the appropriate configuration parameters and controls the test receiver. Furthermore, this software allows the results to be saved while the measurement is running. The test receiver settings for measuring Gaussian and impulsive noise are detailed below.

The Gaussian noise component can be measured in two different ways. On the one hand, an r.m.s. detector can be used. It forms the r.m.s. value of the measured values within a pixel. Alternatively, measurements can be collected in the time domain using a sample detector. Then the r.m.s value must be calculated as ITU-R Report SM.2055 [14] indicates.

$$V_{r.m.s.} = \sqrt{\frac{1}{N}\sum_{i=1}^{N} v_i^2} \qquad (1)$$

where:

$N$ is the number of samples

$v_i$ is the measured sample

$i = 1...N$

When measuring impulsive noise, fast sampling is required in order to detect all the pulses. The right choice is to use a sample detector in the time domain. The sample detector displays the instantaneous value of the level at a pixel, so the sampled data are shown without any further evaluation. The test receiver settings used for all impulsive noise measurements are listed below.
- Mode: spectrum analyzer.
- Detector: sample.
- Span: zero span. For the time domain.

Further details about the receiver are available in [2].

Recommendations ITU-R P.372 [12] and ITU-R SM.1753 [2] explain the antenna specifications to measure noise. The dipole is the recommended antenna for frequencies above 30 MHz. For frequencies below 30 MHz, it is advised to utilize a short vertical monopole as an antenna, since a vertical dipole is too long. For noise measurements, active elements are not recommended [15].

### 4.2 Frequency Selection

The frequency band selected for noise measurements must be free of emissions. The best way to check that there are not emissions is to explore the band of interest. The purpose is to choose a frequency band containing only Gaussian noise. Impulsive noise component must not appear. It is recommended to examine the frequencies of interest many times before taking noise measurements [2].

### 4.3 Procedure for Collecting Data

Once the frequencies have been selected, the impulsive noise source and the location have to be chosen. First of all, it is advisable to check that the level of WGN is low, since it can change from one place to another. Moreover, there is a need to ensure that impulsive noise is only generated by the studied source. The procedure to be followed to take proper measurements is described below.

Firstly, the noise level in the location has to be determined. For this purpose, the WGN measurement is needed, which is obtained making a measurement when the source of interest is not working. A measurement for around two or three minutes per each frequency is required. It is used to evaluate the r.m.s. level in that location and therefore, the Gaussian noise.

After examining the previous measurements it is necessary to verify that impulsive noise is not present. As a practical rule, it can be checked that there are not levels exceeding 13 dB above the r.m.s. value.

Then, the impulsive noise measurements are taken. As mentioned above, the impulses are associated with changes in the operation of the device called events, so before measuring, the event that is going to be studied has to be chosen. To ensure that impulsive noise is generated by the studied source, it is advisable to verify that impulses appear when changing the operational status of the device. For each frequency, more than one measurement studying the same event is needed, in order to estimate average values at a later time.

Figure 3 shows the difference between a measurement that contains Gaussian noise and a measurement in which both components can be distinguished, Gaussian and impulsive. The two measurements have been taken in the same scenario, at 1620 kHz. The difference is that a specific source of impulsive noise was working when taking the second one. The noise was caused by turning on eight fluorescent tubes at the same time. The location where these measurements have been taken is a classroom of the Faculty of Engineering of the University of The Basque Country, where there was no device working, apart from the source that was studied. The sample detector has been used in both measurements, collecting 8001 samples per second. The duration of each measurement is 4 seconds, which corresponds to 32004 samples.

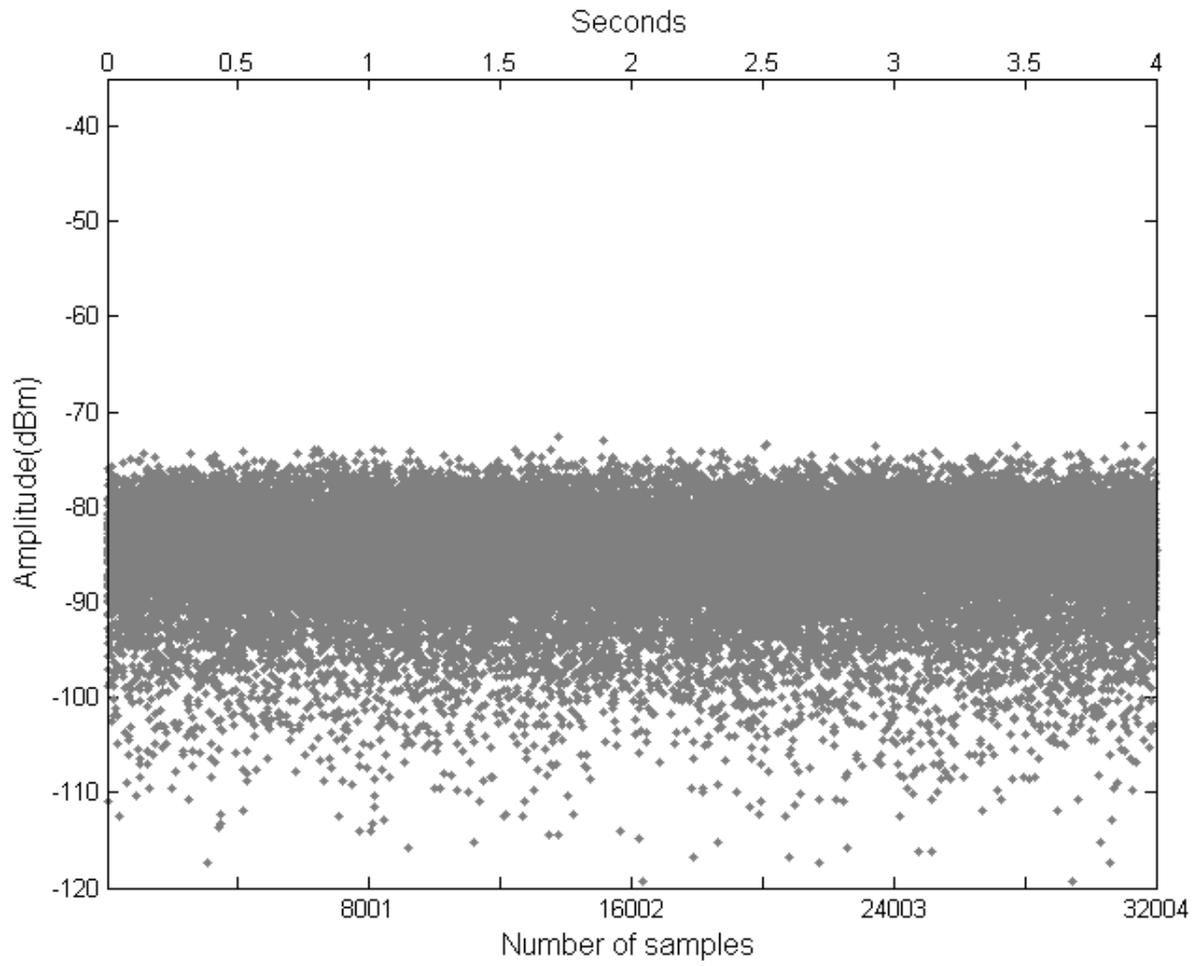

Figure 3. Results of the measurements: (a) WGN measurement

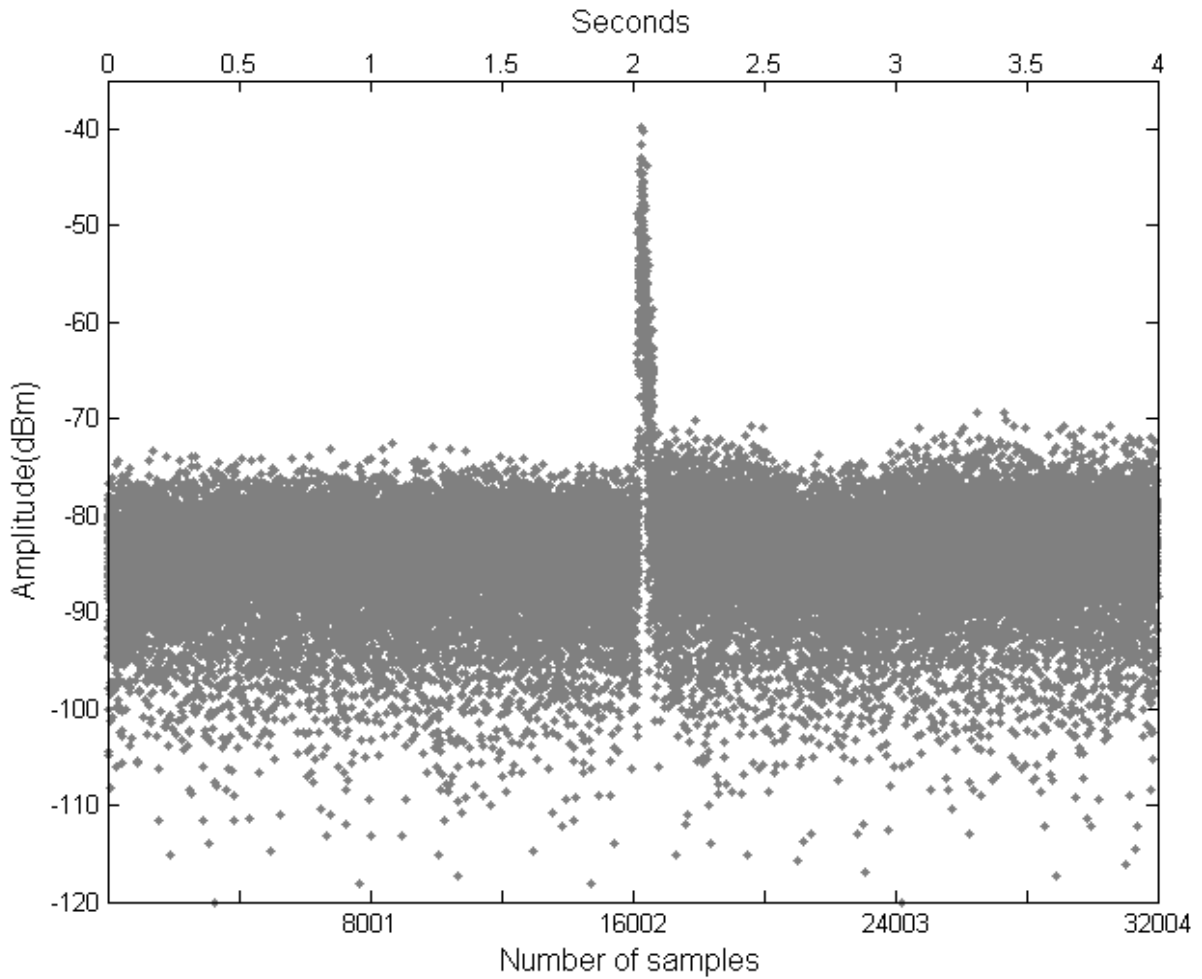

Figure 3. Results of the measurements: (b) IN measurement.

Figure 3. Results of the measurements: (a) WGN measurement, (b) IN measurement.

## 5. Data Processing

Noise samples are obtained as a result of the measurements, but it is necessary to analyse these samples so as to characterize impulsive noise generated by a main source. This procedure is based on ITU-R SM.1753 [2], but some contributions have also been added. The functions that this procedure develops are listed below.

- Separate impulsive noise samples from Gaussian noise
- Combine pulses to bursts
- Determine the impulsive noise present in a measurement
- Characterize the impulsive noise generated by a main source

**5.1 Separate Impulsive Noise Samples from Gaussian Noise**

To separate the impulsive noise samples, the level of Gaussian noise must be determined for each location and each frequency band. For this purpose, the measurements made when the main impulsive source is not working are necessary. The r.m.s. value of all the samples present in a WGN measurement is calculated. Then a threshold is set to 13 dB above the r.m.s. value. After that, IN measurements are analyzed. Samples considered as impulses are those which are above the threshold. The reason for setting the threshold 13 dB above the r.m.s. is that 13 dB is the crest factor for WGN, that is the difference between r.m.s. and peak value. In this way, only samples originating from impulsive noise are extracted [1]. Figure 4 shows the separation between the samples belonging to impulses and the samples associated with Gaussian noise

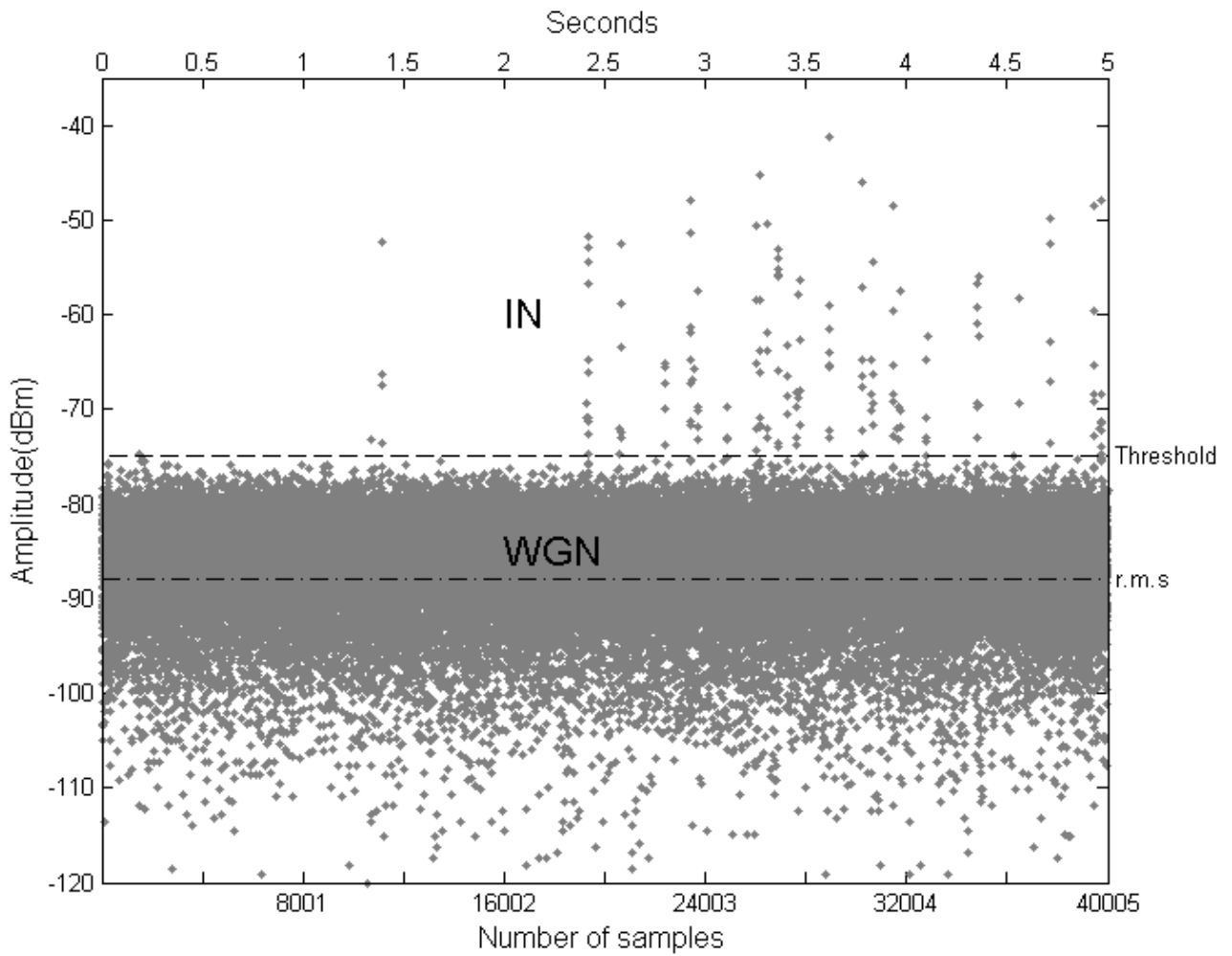

Figure 4. Separation of IN from WGN

### 5.2 Combine Pulses to Bursts

Up to now, samples belonging to impulses have been detected. If the amplitude of measured samples versus time is examined, it can be seen that a series of short peaks or pulse trains are very close to each other. It is necessary to integrate these peaks to a burst. Recommendation ITU-R SM.1753 [2] provides the requirements to combine IN samples to bursts and gives the steps to calculate the duration of each burst in a record.

Combining pulses to bursts in such a way ensures that more than 50% of all samples inside each burst are above the threshold. Consequently, some samples below the threshold can be part of a burst, if they satisfy some conditions. Each burst is characterized by the following parameters:

- Burst duration is the time difference between the first and the last sample of a burst.
- Burst amplitude is the linear average of all samples belonging to a burst, regardless of whether they are above or below the threshold.

In Figure 5 some examples of pulses that are combined to create bursts are shown. A measurement of the noise caused when turning on eight fluorescent tubes at the same time is represented in Figure 5 (a). This measurement has been taken at 630 kHz. The sample detector has been used, collecting 8001 samples per second. As a result, four bursts are obtained which are illustrated in Figure 5 (b).

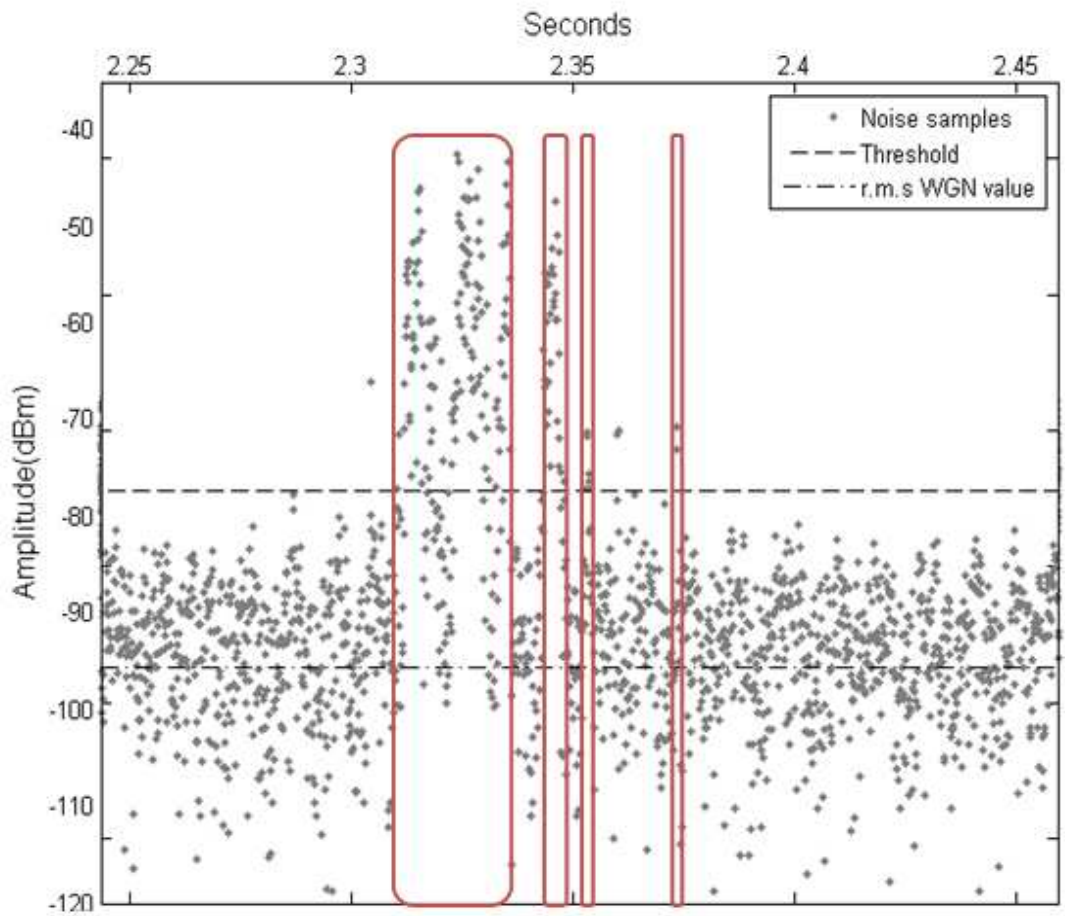

Figure 5. Combining pulses to bursts: (a) IN measurement.

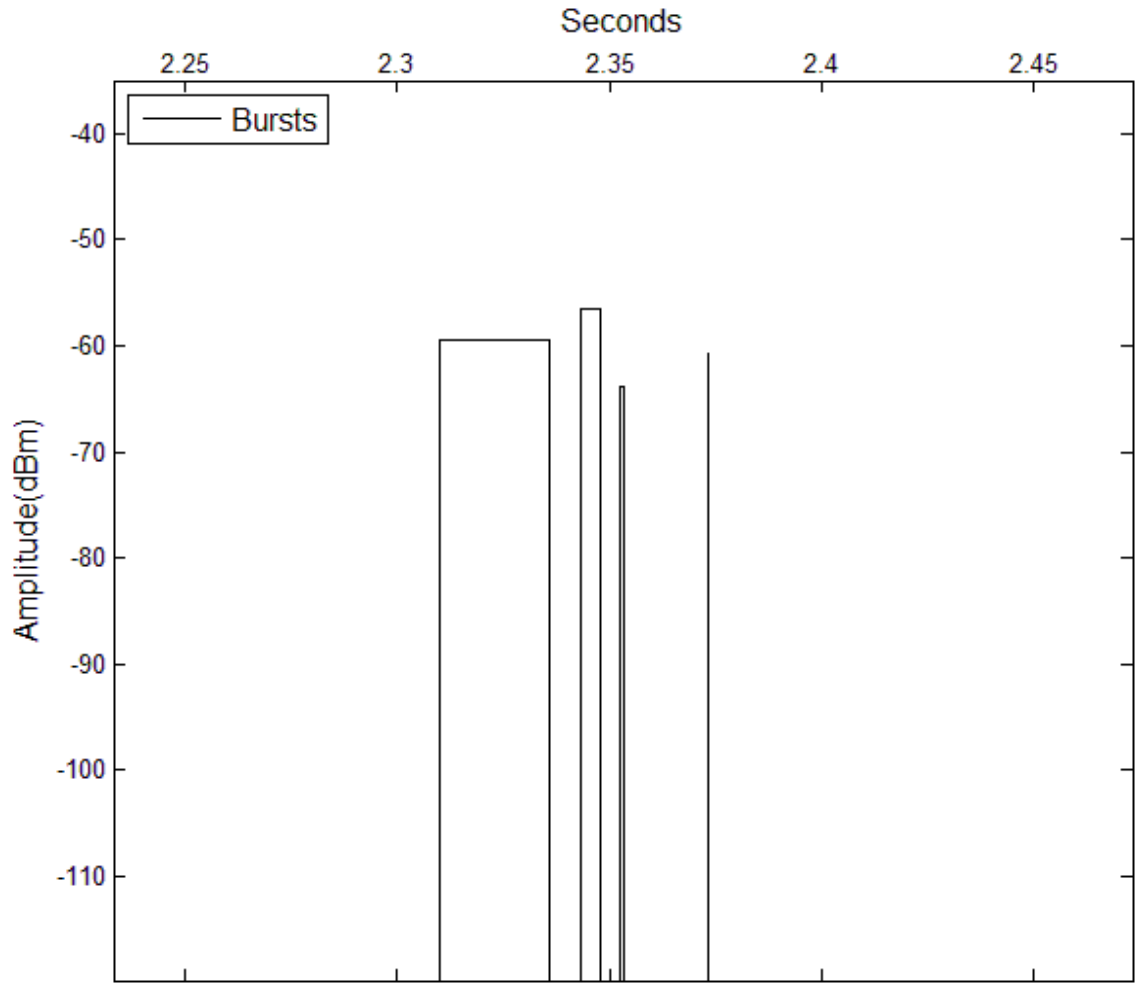

Figure 5.  Combining pulses to bursts (b) Result presentation.

Figure 5.  Combining pulses to bursts: (a) IN measurement, (b) Result presentation.

**5.3 Determine the Impulsive Noise Present in a Measurement**

After processing an impulsive noise measurement, more than one burst can appear. In these cases, the average values of burst parameters must be calculated considering all the bursts that appear in the measurement, as long as these bursts are associated to one particular event. The parameters that describe the impulsive noise detected in a measurement are listed below.

- Number of bursts is the number of the bursts that appear in a measurement.
- Average burst duration is the average duration of the bursts that appear in a measurement. It is calculated as the linear average of all the burst durations existing in a measurement.

- Average burst amplitude is the weighted average amplitude, taking into account all the burst amplitudes existing in a measurement. The longer bursts contribute more than others.
- Average burst separation. This parameter exists only when more than one burst appears. It is calculated as the linear average of all the burst separations existing in a measurement. A burst separation is the distance between two consecutive bursts.

An example of the calculation of IN parameters, considering one measurement, is detailed below:

Number of bursts = N

$$Average\ burst\ duration = \frac{1}{N}\sum_{i=1}^{N} Dur_i \quad (2)$$

$$Average\ burst\ amplitude = \frac{\sum_{i=1}^{N} Amp_i\ Dur_i}{\sum_{i=1}^{N} Dur_i} \quad (3)$$

$$Average\ burst\ separation = \frac{1}{N-1}\sum_{i=1}^{N-1} Sep_i \quad (4)$$

where:
$Dur_i$ is the Burst Duration of the *ith* burst
$Amp_i$ is the Burst Amplitude of the *ith* burst
$Sep_i$ is the Burst Separation between the *ith* burst and the *ith+1* burst

Sometimes, during an impulsive noise measurement, a burst is much longer than the rest. So, analyzing this main burst separately from the others can be useful.

**5.4 Characterize the Impulsive Noise Generated by a Main Source**

Once the measurements of the noise generated by a specific source have been analyzed separately, average values of the parameters must be calculated, so as to fully characterize the impulsive noise source. All the measurements studying the same event at a frequency are necessary to estimate the following parameters:

- Number of bursts is the linear average of the number of bursts considering all measurements.
- Average burst duration is the linear average of the average burst duration of all measurements.
- Average burst amplitude is the linear average of the average burst amplitude of all measurements.
- Average burst separation is the linear average of the average burst separation of all measurements

Apart from average values of the parameters, it is advisable to show how much variation from the mean exists. Therefore, standard deviation of the average burst duration, burst amplitude and burst separation must be given.

For a series of *n* measurements of the same event, the standard deviation characterizes the dispersion of the results. It is calculated as follows:

$$s = \sqrt{\frac{\sum_{j=1}^{n}(q_j - \bar{q})^2}{n-1}} \qquad (5)$$

where:

$q_j$ is the result of the *jth* measurement

$\bar{q}$ is the linear average of the *n* results considered

## 6. Result Presentation

An example of two IN measurements studying the same event, and the bursts representation after data processing is shown in Figure 6. The noise source studied in this example is the fluorescent light. Concretely, the noise is caused when turning on seven fluorescent tubes which flicker for a few seconds before turning on completely. The duration of each measurement is 5 seconds, which corresponds to 40005 samples, since 8001 samples per second have been collected. These measurements have been taken at a frequency of 1910 kHz.

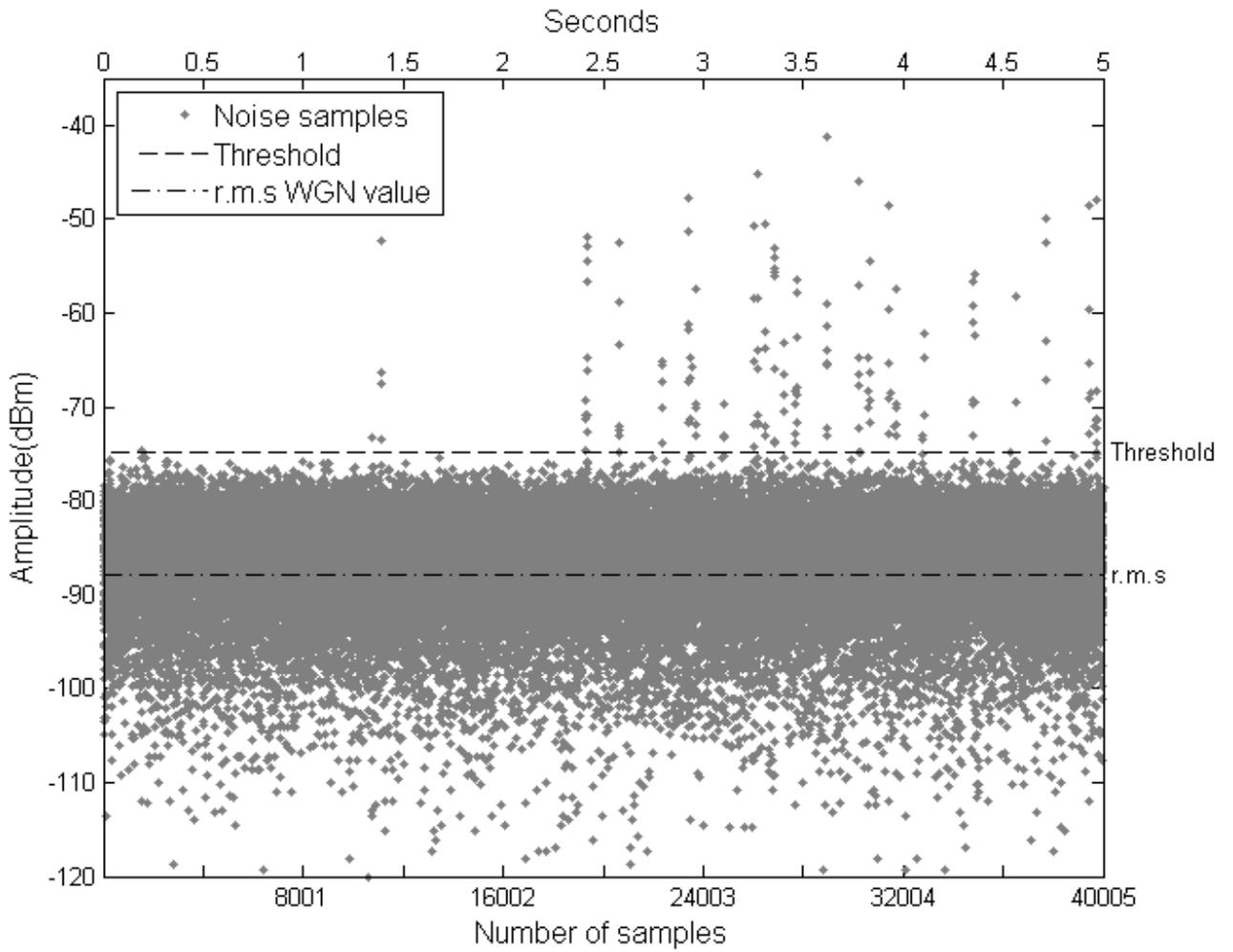

Figure 6. Two IN measurements and its result presentation: (6.1.1) IN measurement 1.

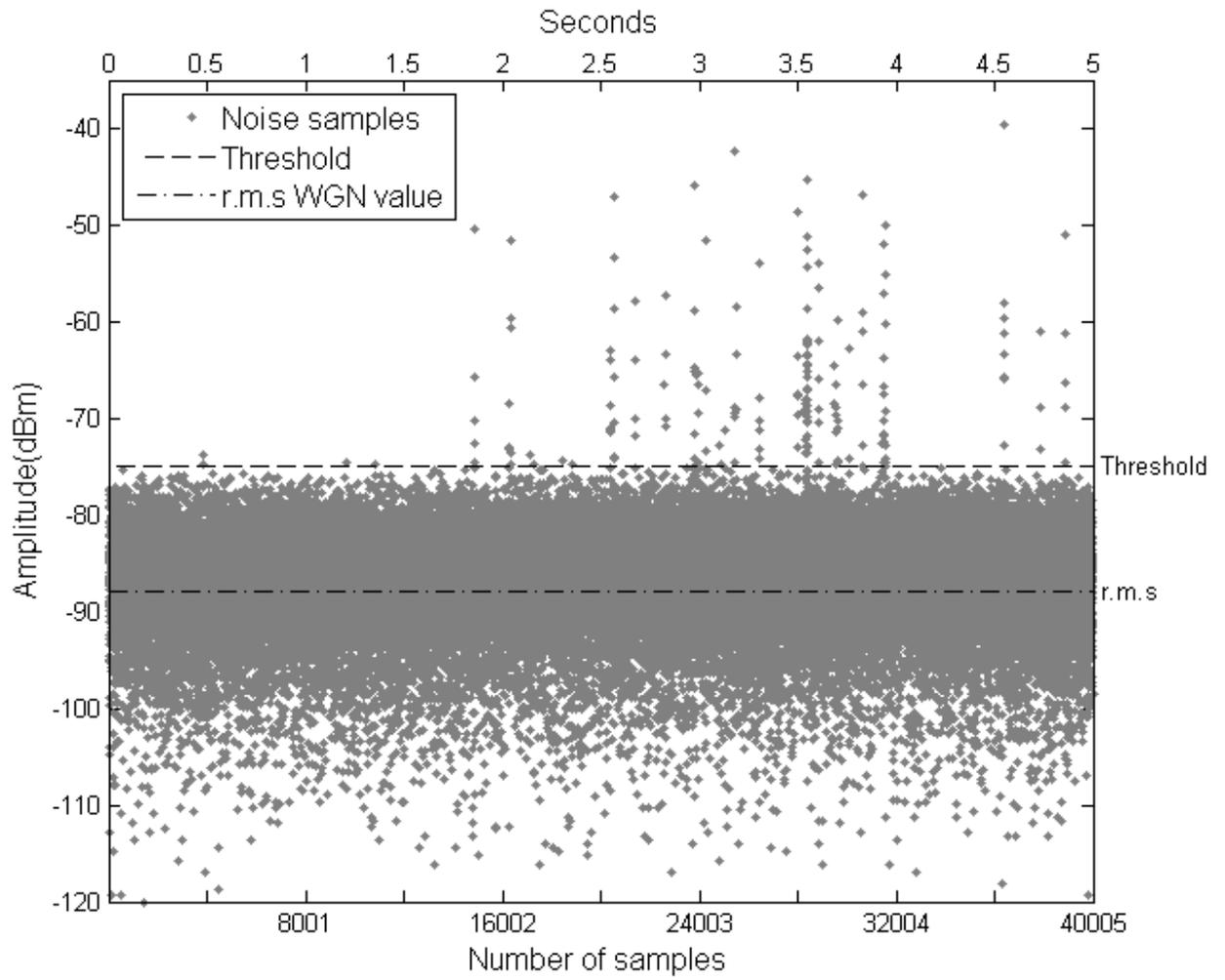

Figure 6. Two IN measurements and its result presentation: (6.1.2) IN measurement 2.

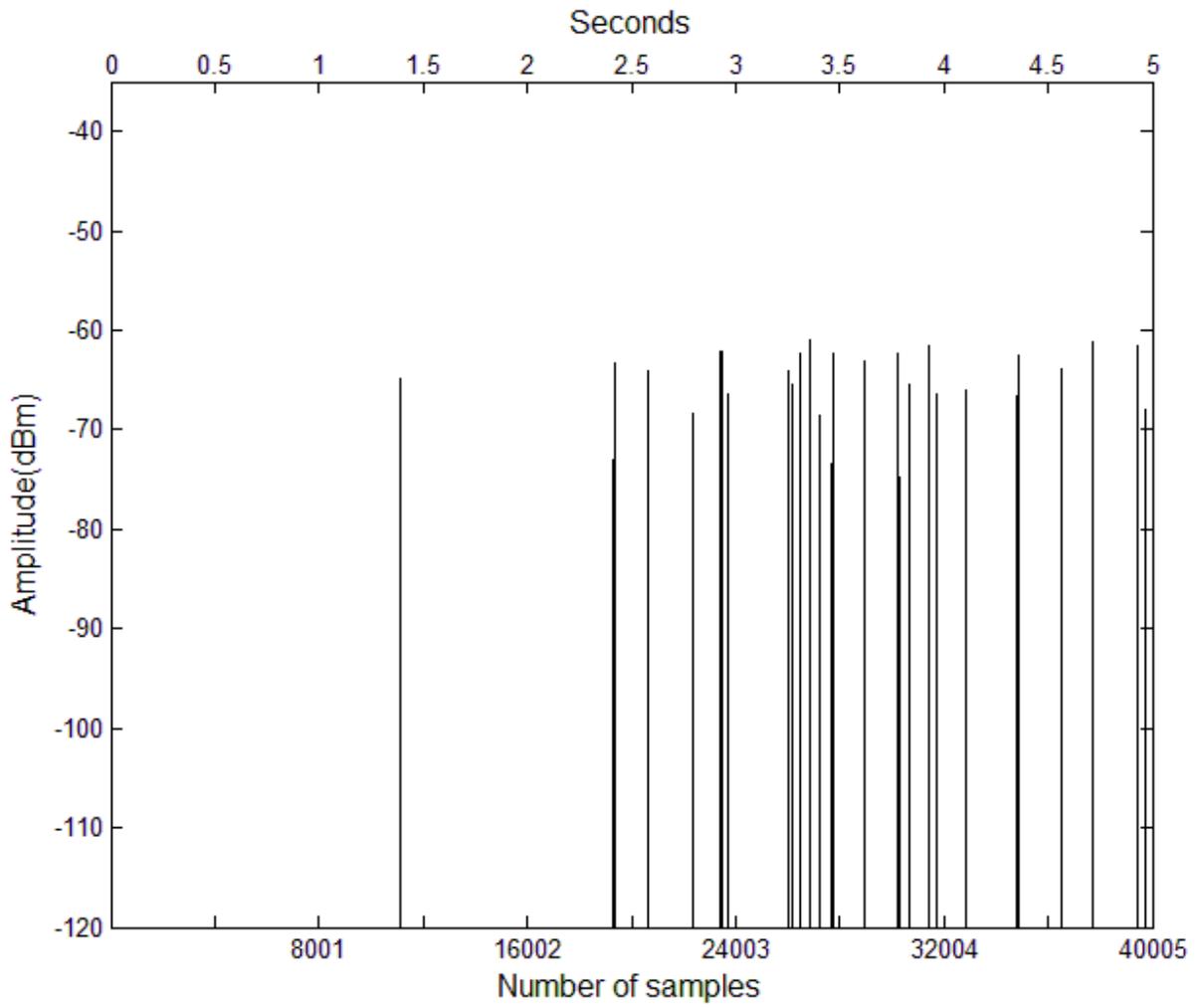

Figure 6. Two IN measurements and its result presentation: (6.2.1) IN measurement 1 result presentation.

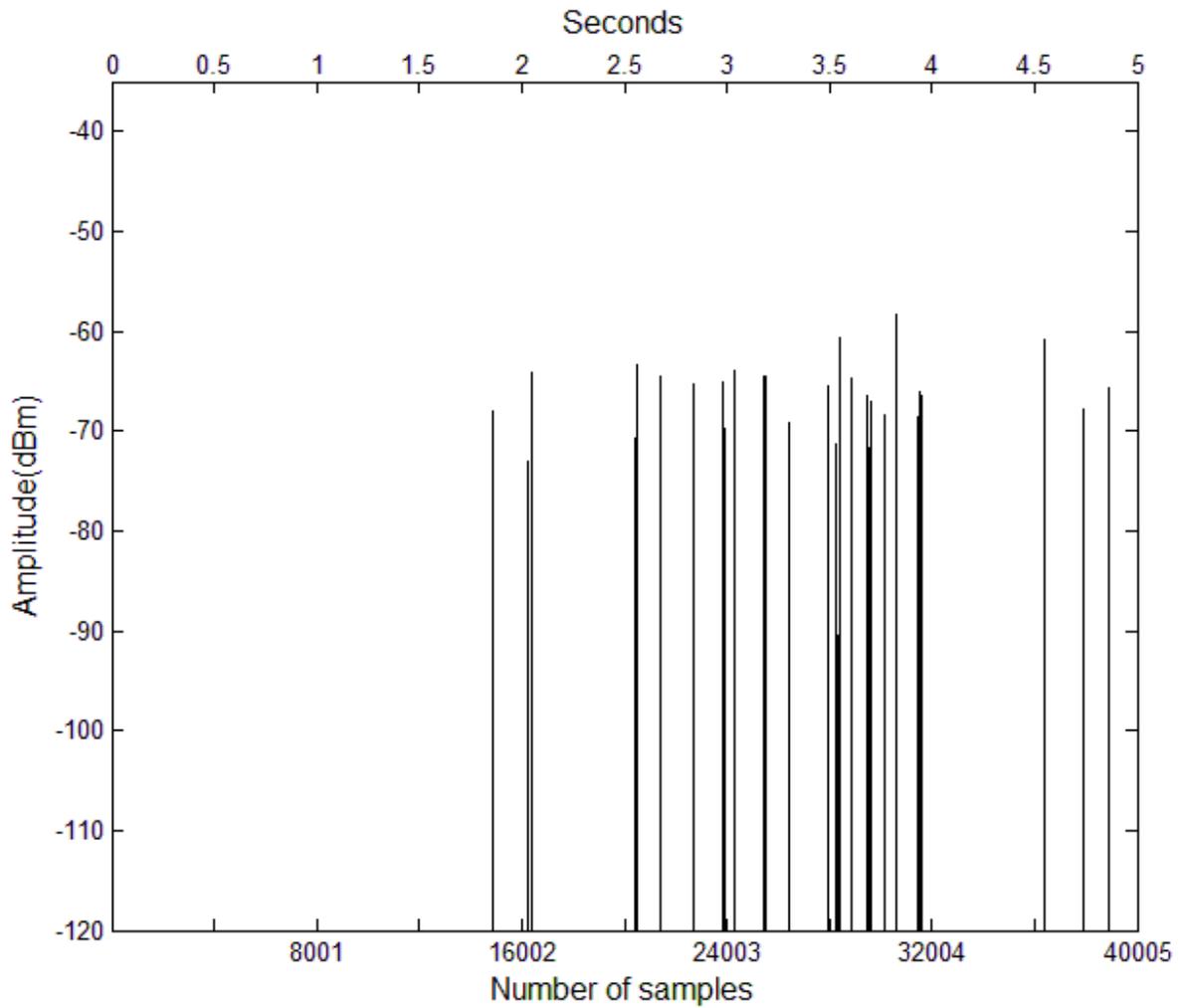

Figure 6. Two IN measurements and its result presentation: (6.2.2) IN measurement 2 result presentation.

Figure 6. Two IN measurements and its result presentation: (6.1.1) IN measurement 1; (6.1.2) IN measurement 2, (6.2.1) IN measurement 1 result presentation (6.2.2) IN measurement 2 result presentation.

Apart from bursts representation, it is useful to give the values of the parameters. So, the parameters that describe the impulsive noise in each measurement are given in Table 1. Then, average values must be calculated in order to characterize the impulsive noise generated, in this case, when turning on seven flickering tubes. These average values are reported in Table 2.

**Table 1. IN parameters of flickering lights turning on event**

| Parameter | Measurement 1 | Measurement 2 |
| --- | --- | --- |

| Number of Bursts | 31 | 30 |
| --- | --- | --- |
| Average Burst Duration (ms) | 0.53 | 0.65 |
| Average Burst Amplitude (dBm) | -64.70 | -66.21 |
| Average Burst Separation (ms) | 118.91 | 103.13 |

**Table 2. Average values of IN parameters of flickering lights turning on event**

| | |
| --- | --- |
| Number of Bursts | 30.5 |
| Average Burst Duration (ms) | 0.59 |
| Standard Deviation of Duration (ms) | 0.08 |
| Average Burst Amplitude (dBm) | -65.46 |
| Standard Deviation of Amplitude (dBm) | 1.07 |
| Average Burst Separation (ms) | 111.02 |
| Standard Deviation of Separation (ms) | 11.16 |

It can be seen that this way of analyzing impulsive noise, only consider the existing noise between the first and the final burst, regardless of how many samples have been taken before impulses detection. Similarly, samples taken after the last detected burst, do not affect the final result.

Previous studies have characterized the impulsive noise using the amplitude probability distribution (APD) [3,16,17]. It consists of a graph which shows the percentage of measurement samples that exceed a certain level or amplitude. In this case, all the samples must be sorted in ascending order, so as to count how many samples exceed that level.

An example of typical amplitude probability distribution is illustrated in Figure 7. Plotting the APD allows separating different components of noise. White Gaussian noise shows up as a straight sloping line and the rising edge to the left indicates impulsive noise.

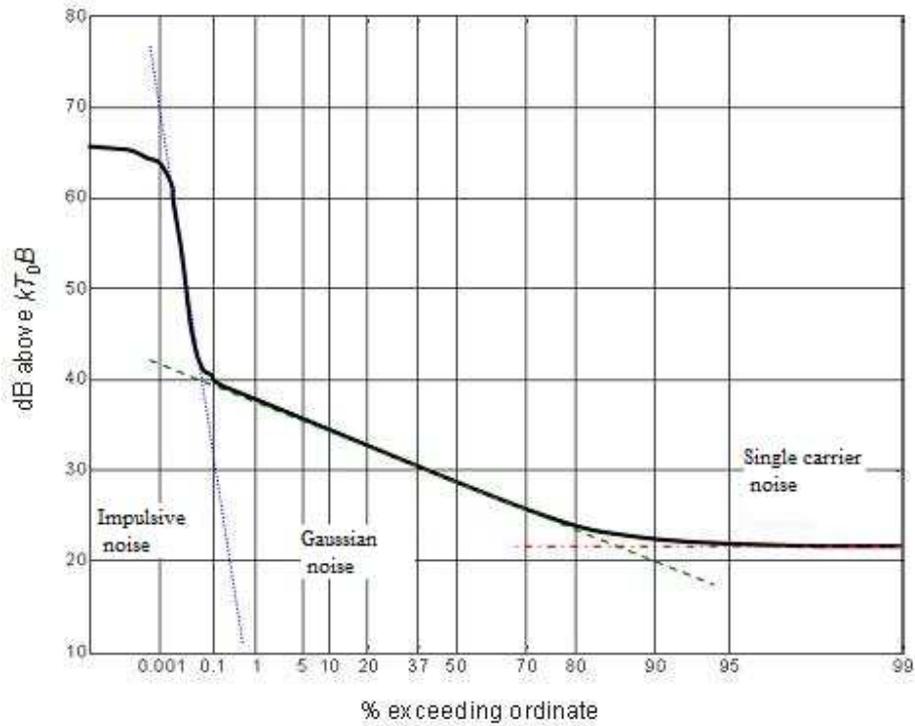

Figure 7. Typical amplitude probability distribution [2].

Figure 8 shows the APD graph corresponding to one of the measurements previously described in Figure 6. Two lines are represented. One corresponds to a WGN measurement and the other one shows the result of the measurement when the specific impulsive noise source is working.

The IN measurement has been taken for 5 seconds, so 40005 samples per measurement have been collected.

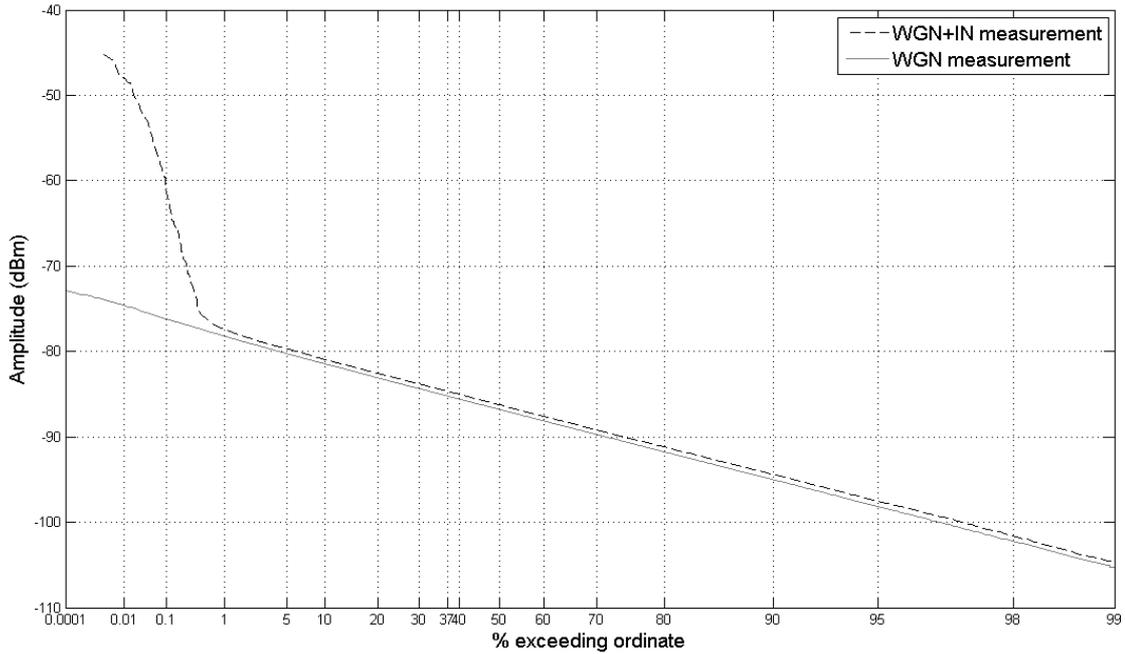

Figure 8. APD graph in the presence of IN.

It is obvious that characterizing impulsive noise using bursts representation and the parameters previously described, gives more information about the pulses than using the amplitude probability distribution. Not only because the parameters that describe the bursts are obtained, but also because the exact position of the pulses is given. In addition, it makes it easier to compare different measurements of impulsive noise.

### 7. Discussion

Impulsive noise measurements are interesting to be studied because it is an important component of radio noise that can significantly interfere with radio services. So it should be taken into account in all radio systems. For this reason, there is a need to establish a methodology for taking and processing noise data, which furthermore allows comparing different measurements.

But there are some difficulties in the data collection and in its later processing. On the one hand, impulsive noise measurements require fast data sampling, because the pulse duration can be very short. If the sampling frequency is not fast enough, some pulses can be lost, so there will be fewer bursts in the final result, or the bursts will be shorter than they should be. On the other hand, data evaluation is complex and requires extensive post processing for several reasons. Firstly, the data volume that needs to be analyzed is much larger than if only Gaussian noise is studied, when taking measurements of the same duration. Another reason is that the parameters that characterize impulsive noise cannot be measured directly. These parameters are determined later, in the data processing, following the procedure indicated in Section 5.

The proposed methodology is based on ITU-R recommendations and according to them, IN samples are those which are at least 13 dB above the WGN. This means that the results of impulsive noise measurements depend on the level of Gaussian Noise. For this reason, it is very important to choose an environment with a noise level as low as possible and naturally, without any impulses apart from those generated by the source of interest. For example, the results presented in this paper have been collected in different parts of the Faculty of Engineering of the University of The Basque Country. Test measurements were carried out in different places of this building in order to select locations with the lowest noise level.

Finally, in order to allow comparing results obtained from different sources, it is important to describe the scenarios in detail.

## 8. Conclusions

This study describes the methodology for measuring and processing impulsive noise when it is generated by a specific source. The instructions to take IN measurements are given, including some specifications about the equipment or the environment. Furthermore, in order to achieve successful results, the steps required in data processing are explained. In the data evaluation, the appropriate parameters to describe the studied noise component are also determined. These parameters make possible the comparison between different measurements.

As a result, impulsive noise sources can be characterized so the noise caused by them can be evaluated.

This work has been made according to the latest recommendation given by the ITU-R SM.1753 [2]. However, in this recommendation there is a lack of information about measuring and evaluating the impulsive noise caused by a specific source. For this reason, some contributions have been added. Among others, the calculation of burst amplitude is defined, since the recommendation only specifies how to calculate burst duration and the separation between bursts. After then, the procedure to calculate average values of the noise present in a measurement is explained. Finally, the procedure to obtain parameters to characterize the impulsive noise generated by a principal source is described.

Furthermore, this work gives answer to the request of the International Telecommunication Union, ITU, in ITU-R 214-4/3 question [18], which considers that it is essential to determine impulsive noise parameters and to obtain reference values.

Previous studies have analyzed impulsive noise component in several ways, but not following the same methodology. A typical way of characterizing that component is to use the amplitude probability distribution, but, as it has been explained before, the methodology used in this study provides more information about the noise.